\documentclass[aps,prl,reprint,superscriptaddress,nofootinbib]{revtex4-1}
\usepackage{graphicx}
\usepackage{subfigure}
\usepackage{amsmath}
\usepackage[section]{placeins}
\usepackage{slashed}
\usepackage{color}
\usepackage{soul}
\usepackage{appendix}
\usepackage{placeins}
\usepackage{float}
\usepackage[utf8]{inputenc}
\usepackage{pstricks}
\usepackage{multirow}
\usepackage{longtable}
\usepackage{hyperref}
\usepackage{subfigure}
\usepackage{epsfig}
\usepackage{braket}
\usepackage[normalem]{ulem} 
\usepackage{longtable,booktabs}
\newcommand{\be}{\begin{equation}}
\newcommand{\ee}{\end{equation}}
\newcommand{\ben}{\begin{eqnarray}}
\newcommand{\een}{\end{eqnarray}}

\usepackage{stackengine}
\usepackage{bbm}
\usepackage{lipsum} % 调用跨栏长公式宏包

\usepackage{verbatim}
\usepackage{bm}
\usepackage{tikz}
\usetikzlibrary{tikzmark}

\begin{document}
%=========================================================================&
%=========================================================================&
\title{Femtoscopic Correlation Functions in Density Operator Representation}

\author{Hao-Nan Liu}
\affiliation{School of Physics, Beihang University, Beijing 102206, China}
\affiliation{Instituto de Física, Universidade de São Paulo, Rua do Matão, São Paulo, SP, 05508-090, Brazil}

\author{Duo-Lun Ge}
\affiliation{School of Physics, Beihang University, Beijing 102206, China}

\author{Zhi-Wei Liu}
\affiliation{Institute for Advanced Study in Nuclear Energy \& Safety, College of Physics and Optoelectronic Engineering, Shenzhen University, Shenzhen 518060, China}
\affiliation{Shenzhen Key Laboratory of Nuclear and Radiation Safety, Shenzhen 518060, China}

\author{Jun-Xu Lu}
\affiliation{School of Physics, Beihang University, Beijing 102206, China}

\author{Li-Sheng Geng}
\email[Corresponding author: ]{lisheng.geng@buaa.edu.cn}
\affiliation{Sino-French Carbon Neutrality Research Center, \'Ecole Centrale de P\'ekin/School of General Engineering, Beihang University, Beijing 100191, China}
\affiliation{School of Physics, Beihang University, Beijing 102206, China}
\affiliation{Peng Huanwu Collaborative Center for Research and Education, Beihang University, Beijing 100191, China}
\affiliation{Southern Center for Nuclear-Science Theory (SCNT), Institute of Modern Physics, Chinese Academy of Sciences, Huizhou 516000, China}

%=========================================================================

\begin{abstract}
Femtoscopic correlation functions (CFs) have been increasingly used to extract strong interactions between pairs of unstable particles, but their physical soundness has recently been questioned. To answer this, we formulate CFs at the operator level, with the observed subsystem described by a reduced density operator and subsequent dynamics absorbed into an effective measurement operator. The Koonin-Pratt form is recovered under four well-motivated reductions. The formulation makes explicit that the source-side and interaction-side representations must be consistently matched, and motivates an operational convention in which a measured reference correlation establishes a compatible source--interaction pairing that can be extended to other pairs for CF-to-CF predictions.
\end{abstract}

\maketitle

\textit{Introduction: }
Femtoscopic correlation functions (CFs) trace their conceptual origin to Hanbury Brown-Twiss (HBT) intensity interferometry, where intensity correlations were used to infer stellar angular sizes~\cite{HanburyBrown:1956bqd}.
CFs probe femtometer-scale source sizes, collective expansion, freeze-out geometry, emission lifetimes, and homogeneity regions~\cite{Boal:1990yh,Wiedemann:1999qn,Weiner:1999th,Lisa:2005dd}.
In a widely used form, measured correlation functions are related to a source function and a final-state scattering wave function through the Koonin-Pratt (K-P) formula~\cite{Koonin:1977fh,Pratt:1990zq}
\begin{equation}\label{Eq.K-P}
C(\mathbf{q})=\int \mathrm{d}^3r\,S(\mathbf{r})\,\left|\psi_{\mathbf{q}}(\mathbf{r})\right|^2,
\end{equation}
which has been highly successful as a phenomenological description of femtoscopic data~\cite{Boal:1990yh,Bauer:1992ffu,Lisa:2005dd,Fabbietti:2020bfg} and as a practical bridge between measured momentum correlations and space-time source information~\cite{Pratt:1990zq,Chapman:1994xa,Csorgo:1995bi,Wiedemann:1999qn,STAR:2004qya,STAR:2009fks,ALICE:2011dyt}.

The same formalism is increasingly used in non-traditional femtoscopy~\cite{Mihaylov:2021glj}, not only to determine the emission source but also to constrain or extract interactions between emitted hadrons~\cite{STAR:2015kha,ALICE:2018ysd,ALICE:2019hdt,Fabbietti:2020bfg,ALICE:2020mfd,Si:2025eou}.
In this setting, the source also serves as an input, and its definition, universality, and consistency with the interaction model directly affect the extracted final-state interaction (FSI)~\cite{Bialas:2005ps,Epelbaum:2025aan}.
The standard K-P expression implicitly assumes a local classical source-side probability density, an effective two-body wave function for the relevant final-state dynamics, and negligible nonlocal source structures.
These assumptions are often reasonable but become nontrivial as femtoscopic extractions of strong interactions become increasingly precise~\cite{Epelbaum:2025aan}.

The present work has two goals. First, we formulate femtoscopic correlation functions systematically at the operator level, treating the observed $n$-particle channel as an explicitly retained subsystem embedded in a larger total system.
The correlation function is then organized as the expectation value of the effective measurement operator with respect to the reduced density operator. In this language, the phase-space emission function is the Wigner representation of the reduced density operator~\cite{Lednicky:2007ax}, while the K-P source function is recovered only after the sudden, equal-time, on-shell, and diagonal/decoherence approximations. 
Second, we clarify off-shell ambiguity and source--interaction matching~\cite{Epelbaum:2025aan}.
The invariance of the correlation function under simultaneous transformations of the source density operator and the scattering state implies that their representations must be matched consistently. In phenomenological applications, however, this matching is not automatic: the source kernel and scattering wave function are usually constructed independently, while on-shell scattering data constrain a scattering-equivalence class rather than a unique off-shell or interior realization of the scattering wave function. The transformations considered in Ref.~\cite{Epelbaum:2025aan} provide an explicit example showing that an uncontrolled mismatch can produce a sizable correlation-level effect. We therefore introduce an operational source--interaction convention in which the measured correlation function of a reference pair is used to calibrate a compatible interaction-side realization and establish an effective source--interaction pairing.

\textit{Theoretical framework: } We begin from the complete production process and describe the full system by a density operator $\hat\rho^{\mathrm{tot}}$ defined on the total Hilbert space
\begin{equation}
\mathcal H_{\mathrm{tot}}=\mathcal H_n\otimes\mathcal H_E,
\end{equation}
where $\mathcal H_n$ and $\mathcal H_E$ denote the Hilbert spaces of the observed $n$-particle subsystem and its environment of unobserved degrees of freedom, respectively.
Tracing over the environment yields the reduced density operator relevant for the measured subsystem of interest
\begin{equation}\label{Eq.TraceofTotalDensityOperator}
\hat\rho^{(n)}_{\mathrm{in}}=\mathrm{Tr}_{\mathcal H_E}\,\hat\rho^{\mathrm{tot}}.
\end{equation}

In the present formulation, the tracing stage need not coincide with the conventional sudden approximation~\cite{Smith:2026ocz}. 
We define ``femtoscopic freeze-out'' as the last-interaction stage where the observed $n$-particle subsystem effectively separates from the environment, after which the mutual $n$-body FSI approximates its dynamics.
If the environment is traced out earlier, additional medium-induced effects, such as interactions between the observed particles and unobserved hadrons, must be included in the subsequent reduced dynamics~\cite{Lin:2002gc,Hirano:2005xf}.
This evolution is then described by an effective open-system map $\mathcal E_{\rm full}$~\cite{breuer2002theory,rivas2012open}. If, instead, the trace is taken at freeze-out and the subsystem is effectively closed afterward, the map reduces to the usual unitary FSI evolution,
\begin{equation}\label{Eq.SuddenApproximation}
\mathcal E_{\rm full}\left[\hat\rho_{\rm in}^{(n)}\right]
\simeq
\hat U_{\rm FSI}\hat\rho_{\rm in}^{(n)} \hat U_{\rm FSI}^{\dagger}.
\end{equation}

In femtoscopy, this limit is the operator-level form of the sudden approximation~\cite{Lednicky:2005tb}. 
Its physical motivation is that, at small relative momenta relevant to femtoscopy, the characteristic time scale of the relative two-particle dynamics is much longer than that of the production process, motivating the separation of post-freeze-out FSI from the production stage~\cite{Lednicky:2005tb}.
Thus, freeze-out is a physically motivated tracing choice rather than a built-in assumption of the present framework.
For generality, we adopt the following effective quantum map $\mathcal E_{\rm full}$ acting on the observed $n$-particle subsystem
\begin{equation}
\hat\rho_{\rm out}^{(n)}
=
\mathcal E_{\rm full}
\left[
\hat\rho_{\rm in}^{(n)}
\right],
\end{equation}
where $\mathcal E_{\rm full}$ is modeled as a completely positive trace-preserving (CPTP) map~\cite{Stinespring:1955eig,choi1975completely}, which admits a Kraus representation~\cite{kraus1971general,kraus1983states}. Such CPTP maps have been employed, for instance, via Lindblad evolution for quarkonium-like states in heavy-ion collisions~\cite{Coci:2026puh}. For notational simplicity, we work in a single-channel scenario in the following.
Extensions to coupled-channel systems are provided in the Supplemental Material (SM)~\cite{SupplementalMaterial}.

For a measurement described by $\hat M_n(p_1,\dots,p_n)$, where $p_i$ is the four-momentum of the $i$-th observed particle, the generalized correlation function is given by
\begin{equation}\label{Eq.ParentFormula_DensityOperatorEvolution}
C_n(p_1,\dots,p_n)
=
\mathrm{Tr}_{\mathcal H_n}
\left[
\mathcal E_{\rm full}\left[\hat\rho_{\rm in}^{(n)}\right]
\hat M_n(p_1,\dots,p_n)
\right].
\end{equation}

Equivalently, using the adjoint map $\hat M_{n}^{\rm eff} = \mathcal E_{\rm full}^{\dagger}\left[\hat M_n\right]$, one obtains
\begin{equation}\label{Eq.ParentFormula}
C_n(p_1,\dots,p_n)
=
\mathrm{Tr}_{\mathcal H_n}
\left[
\hat\rho_{\rm in}^{(n)}
\hat M_{n}^{\rm eff}(p_1,\dots,p_n)
\right],
\end{equation}
which can be easily proven by using the Kraus form $\mathcal E_{\rm full}\left[\hat\rho_{\rm in}^{(n)}\right]=\sum_a \hat{K}_a\hat\rho_{\rm in}^{(n)} \hat{K}_a^\dagger$. Strictly speaking, Eq.~\eqref{Eq.ParentFormula} gives the measured $n$-particle yield associated with the effective measurement operator $\hat M_{n}^{\rm eff}$, rather than a normalized correlation function. For the present discussion, the normalization denominator is kept implicit hereafter.

Eq.~\eqref{Eq.ParentFormula} is the density-operator--measurement form of the generalized correlation function. Up to and including this equation, the construction remains formal: none of the sudden, equal-time, on-shell, or diagonal/decoherence approximations has yet been imposed. The framework can therefore be extended to a broad range of scenarios.

We now turn our focus to the two-particle case. The bare operator for measuring two particles of asymptotic free momenta $p_1$ and $p_2$ is
\begin{equation}\label{Eq.BareMeasurementProjector}
    \hat M_2(p_1,p_2)
    =
    \ket{p_1,p_2}\bra{p_1,p_2},
\end{equation}
which specifies only the choice of the measured final state. To arrive at the conventional K-P formula, we now introduce the \textbf{sudden approximation}. According to Eq.~\eqref{Eq.SuddenApproximation} and Eq.~\eqref{Eq.ParentFormula}, the two-body FSI evolution can be absorbed into the effective measurement operator as
\begin{equation}
\hat M_{2}^{\rm eff}(p_1,p_2)
=
\mathcal E_{\rm full}^{\dagger}[\hat M_2]
\simeq
\hat U_{\rm FSI}^{(2)\dagger}
\ket{p_1,p_2}\bra{p_1,p_2}
\hat U_{\rm FSI}^{(2)}.
\end{equation}

Defining the FSI-modified scattering state by
\begin{equation}
\ket{p_1,p_2;(-)}
=
\hat U_{\rm FSI}^{(2)\dagger}
\ket{p_1,p_2},
\end{equation}
one obtains
\begin{equation}\label{Eq.ParentFormula_with_SuddenApproximation}
C(p_1,p_2)
=
\bra{p_1,p_2;(-)}
\hat\rho_{\rm in}^{(2)}
\ket{p_1,p_2;(-)},
\end{equation}
where $\hat\rho_{\rm in}^{(2)}$ should be understood as the reduced two-body density operator at freeze-out. 

To make the reduction to the relative-coordinate representation explicit, we first insert complete coordinate states $\mathbbm 1 =\int \mathrm d^4x_1\,\mathrm d^4x_2\, \ket{x_1,x_2}\bra{x_1,x_2}$ \footnote{Strictly speaking, four-coordinate states are not an orthonormal complete basis of the physical relativistic Hilbert space, where states are restricted to the positive-energy mass shell. The insertion used here should therefore be understood in an enlarged off-shell coordinate representation, with the final observable obtained only after the on-shell projection. In this sense, the present formulation naturally tracks possible off-shell information.} into Eq.~\eqref{Eq.ParentFormula_with_SuddenApproximation}, and change from $(x_1,x_2)$ and $(p_1,p_2)$ to the center-of-mass and relative coordinates $(X,r)$ and $(P^\mu,q^\mu)$, respectively. In the translationally invariant two-body realization adopted here, the FSI-modified Bethe-Salpeter amplitude $\Phi_{P^\mu,q^\mu}^{(-)}(X,r) \equiv \bra{X,r} \hat U_{\rm FSI}^{(2)\dagger} \ket{P^\mu,q^\mu}$ factorizes into a center-of-mass plane wave and a relative-coordinate part. The center-of-mass dependence can then be absorbed into a fixed-$P^\mu$ relative density matrix\footnote{Eq.~\eqref{Eq.FourDimensionDensityMatrix} defines the reduced density operator in a fixed total-momentum sector $P^\mu$, acting on the full relative-motion Hilbert space. Hence, $q^\mu$, which labels the asymptotic state selected by the measurement, is not an independent label of the density operator; on the physical mass shell, however, the total-momentum label should be understood as $P^\mu=P^\mu(q^\mu)$, as discussed below.},
\begin{equation}\label{Eq.FourDimensionDensityMatrix}
\rho_{P^\mu}(r,r')
\equiv
\int \mathrm d^4X\,\mathrm d^4X'\,
e^{iK\cdot(X'-X)}
\rho_{\rm in}^{(2)}(X,r;X',r'),
\end{equation}
where $K^\mu=p_1^\mu+p_2^\mu=2P^\mu$. This gives the four-dimensional relative-coordinate form
\begin{equation}
C(q^\mu,P^\mu)
=
\int \mathrm d^4r\,\mathrm d^4r'\,
\rho_{P^\mu}(r,r')
\Phi_{P^\mu,q^\mu}^{(-)}(r')\,
\Phi_{P^\mu,q^\mu}^{(-)*}(r).
\end{equation}
Here, $\Phi_{P^\mu,q^\mu}^{(-)}(r)$ denotes the relative-motion Bethe–Salpeter amplitude after the free center-of-mass plane wave has been factored out. The product of the center-of-mass phases then generates $e^{iK\cdot(X'-X)}$, which is absorbed into the relative density matrix in Eq.~\eqref{Eq.FourDimensionDensityMatrix}. However, $\Phi_{P^\mu,q^\mu}^{(-)}(r)$ still depends on the relative time coordinate $t^*$ in the pair rest frame (PRF). Following Ref.~\cite{Lednicky:2005tb}, the non-equal-time effect can be neglected when $|t^*|\ll m(t^*)r^{*2}$, with $m(t^*>0)=m_2$ and $m(t^*<0)=m_1$. Under this condition, we introduce the \textbf{equal-time approximation} (ETA) in the pair rest frame to keep only the equal-time component of the Bethe-Salpeter amplitude
\begin{equation}
\Phi_{P^\mu,q^\mu}^{(-)}(r^0,\mathbf r)
\xrightarrow{\rm ETA}
\Phi_{P^\mu,q^\mu}^{(-)}(0,\mathbf r).
\end{equation}

Correspondingly, the relative-time dependence of the density matrix is absorbed into the equal-time three-dimensional density matrix
\begin{equation}
\rho^{\rm ETA}_{P^\mu}(\mathbf r,\mathbf r')
\equiv
\int \mathrm dr^0\,\mathrm dr^{\prime 0}\,
\rho_{P^\mu}\!\left((r^0,\mathbf r),(r^{\prime 0},\mathbf r')\right).
\end{equation}

After the ETA, one must still specify the kinematics carried by the labels $P^\mu$ and $q^\mu$. Since the detected particles are asymptotic physical states, their four-momenta are on-shell. For simplicity, we present the kinematical constraints for equal masses~\footnote{The unequal-mass case can be treated with the standard convention $p_1^\mu=(1+\alpha)P^\mu+q^\mu$, $p_2^\mu=(1-\alpha)P^\mu-q^\mu$, with $\alpha=(m_1^2-m_2^2)/(4P^2)$~\cite{Smith:2026ocz}.}
\begin{equation}\label{Eq.OnShellConstraint}
(P+q)^2=(P-q)^2=m^2
\,\Rightarrow\,
P\!\cdot q=0,\, P^2+q^2=m^2.
\end{equation}

In the true pair rest frame, $P^{*\mu}=(P^{*0},\mathbf 0)$, and therefore $P^*\!\cdot q^*=0$ gives $q^{*0}=0$. Hence $\Phi_{P^{*\mu},q^{*\mu}}^{(-)}(0,\mathbf r^{*})$ may be denoted as $\Phi_{P^{*\mu},\mathbf q^{*}}^{(-)}(0,\mathbf r^{*})$. In the low-$\mathbf q^{*}$ region, the equal-time Bethe-Salpeter amplitude can be further reduced to the nonrelativistic Schrödinger scattering wave function, i.e., $\Phi_{P^{*\mu},\mathbf q^{*}}^{(-),{\rm ETA}}(\mathbf r^{*}) \simeq \psi_{\mathbf q^{*}}^{(-)}(\mathbf r^{*})$.

The preceding steps reduce the relative dynamics to a three-dimensional wave-function problem, but the source-side label has not yet been simplified. Exact on-shell kinematics requires the pair momentum to depend on the relative momentum, $P^\mu_{\rm lab}=P^\mu_{\rm lab}(\mathbf q^{*})$. For fixed $\mathbf P_{\rm lab}$, Eq.~\eqref{Eq.OnShellConstraint} yields
\begin{equation}
P^0_{\rm lab}(\mathbf q^{*})=\sqrt{m^2+\mathbf q^{*2}+\mathbf P_{\rm lab}^2},
\end{equation}
where $\mathbf q^*$ denotes the relative three-momentum defined in PRF. For brevity, unless stated otherwise, $q$ is defined in the pair rest frame, while $P$ is defined in the laboratory frame; the superscript ``$*$'' and subscript ``lab'' are therefore omitted. Thus different values of $\mathbf q$ correspond to slightly different true PRFs and to a family of density matrices $\rho^{\rm ETA}_{P^\mu(\mathbf q)}(\mathbf r,\mathbf r')$. Equivalently, if one wants to use a source-side object defined only with physical on-shell single-particle momentum labels, keeping the exact $P^\mu(\mathbf q)$ would require an off-shell continuation of that object away from the $\mathbf q=0$ point. To avoid this, we adopt the \textbf{on-shell approximation} to replace the $q$-dependent pair momentum $P^\mu(\mathbf q)$ by the fixed pseudo-pair momentum $p^\mu\equiv P^\mu(\mathbf q=0)$, so that the same pseudo-pair-rest-frame label is used for all relative momenta.

With the above reductions, Eq.~\eqref{Eq.ParentFormula_with_SuddenApproximation} becomes
\begin{equation}\label{Eq.GeneralizedRelativeDensityCF}
C(\mathbf q,\mathbf P)
=
\int \mathrm d^3r\,\mathrm d^3r'\,
\rho_{\mathbf P}(\mathbf r,\mathbf r')\,
\psi_{\mathbf q}^{(-)}(\mathbf r')\,
\psi_{\mathbf q}^{(-)*}(\mathbf r),
\end{equation}
where $\rho_{\mathbf P}$ denotes the fixed-$p^\mu$ equal-time relative density matrix. Eq.~\eqref{Eq.GeneralizedRelativeDensityCF} is the generalized two-body correlation function in the relative-coordinate representation, which keeps the off-diagonal structure of the density matrix. To make the connection with the source-operator language explicit, we rewrite Eq.~\eqref{Eq.GeneralizedRelativeDensityCF} as
\begin{equation}\label{Eq.CFbraketform}
C(\mathbf q,\mathbf P)=\bra{\psi_{\mathbf q}^{(-)}}\hat\rho_{\mathbf P}\ket{\psi_{\mathbf q}^{(-)}},
\end{equation}
where $\hat\rho_{\mathbf P}=\int \mathrm{d}^3r\mathrm{d}^3r^\prime\,\rho_{\mathbf P}(\mathbf r,\mathbf r')\ket{\mathbf r}\bra{\mathbf r'}$. In this notation, the source operator introduced in Ref.~\cite{Epelbaum:2025aan} is naturally identified with the reduced two-body density operator after the same sudden, equal-time, and on-shell reductions, up to the conventional normalization of the correlation function. To recover the Koonin-Pratt formula, one must impose the \textbf{diagonal}, or \textbf{decoherence}, \textbf{approximation}
\begin{equation}\label{Eq.DiagonalApproximation}
\hat\rho_{\mathbf P} \simeq \int \mathrm{d}^3r\,S_{\mathbf P}(\mathbf r)\ket{\mathbf r}\bra{\mathbf r}.
\end{equation}
This approximation is the relative-coordinate counterpart of the traditional smoothness approximation~\cite{Pratt:1997pw}, as discussed in detail in the SM. Under this approximation, and by further assuming that the source function is independent of the average pair momentum $\mathbf P$, Eq.~\eqref{Eq.GeneralizedRelativeDensityCF} reduces to the widely used Koonin-Pratt formula in Eq.~\eqref{Eq.K-P}.

The decoherence interpretation of this approximation can be motivated by viewing the observed pair as an open subsystem embedded in a larger system. After tracing over the unobserved environmental degrees of freedom, the relative-coordinate coherence of its reduced state may be strongly suppressed\footnote{For a recent illustrative study of medium-induced decoherence in heavy-ion collisions within an open-quantum-system framework, see Ref.~\cite{Coci:2026puh}.}. If the remaining off-diagonal information is irrelevant to the observable, the reduced pair may be interpreted as an effective statistical ensemble over relative positions $\mathbf r$, with $S_{\mathbf P}(\mathbf r)$ as its distribution; the correlation is then the ensemble average of $|\psi_{\mathbf q}^{(-)}(\mathbf r)|^2$. Thus, decoherence provides a possible physical realization of the diagonal approximation and gives the traditional source function an effective classical-ensemble interpretation.

The density-operator formulation also clarifies source universality and off-shell ambiguity.
As emphasized in Ref.~\cite{Epelbaum:2025aan}, when the interaction-side representation is changed by a scattering-equivalent unitary transformation $\hat U_S$, the source density operator must be transformed accordingly,
\begin{equation}\label{Eq.SimultaneousTransformation}
|\psi_{\mathbf q}^{(-)}\rangle
\rightarrow
\hat U_S|\psi_{\mathbf q}^{(-)}\rangle,
\qquad
\hat\rho_{\mathbf P}
\rightarrow
\hat U_S\hat\rho_{\mathbf P}\hat U_S^\dagger.
\end{equation}
The simultaneous transformation leaves $C(\mathbf q,\mathbf P)$ in Eq.~\eqref{Eq.CFbraketform} unchanged, showing that the source-side and interaction-side representations must remain consistently matched. As discussed in the SM, at the representation level this simultaneous transformation can equivalently be absorbed into the choice of basis in which the source density operator and the scattering state are represented.

If the source density operator and the scattering state are derived consistently within the same microscopic framework, this matching is automatic. In phenomenological applications, however, the source kernel and scattering wave function are usually constructed independently. Although both may be written using the same relative-coordinate label $\mathbf r$, this does not by itself ensure that they are represented in the same relative-coordinate basis. Directly combining such independently specified representations may therefore generate a source--interaction mismatch. On-shell scattering data alone cannot resolve this matching problem, since they constrain a scattering-equivalence class rather than a unique off-shell or interior realization.

\begin{figure*}[htbp]
\centering
\includegraphics[width=0.95\textwidth]{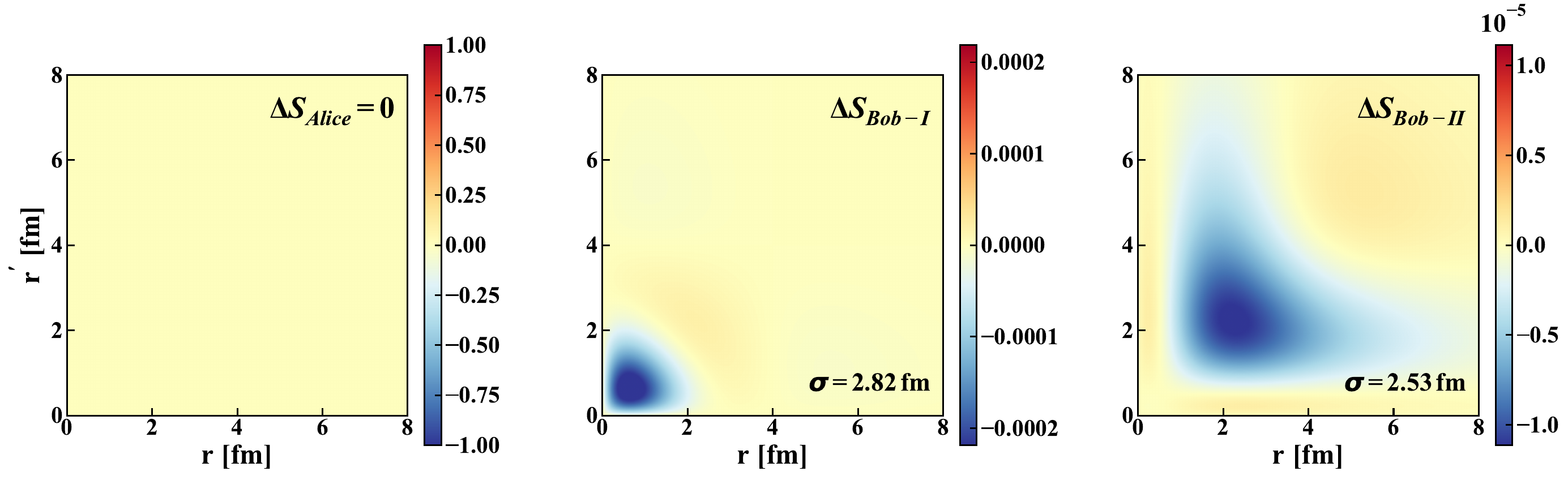}
\caption{
Heat maps of the induced source-kernel change $\Delta S_U(\mathbf r,\mathbf r')$ resulting from absorbing the source--interaction mismatch of Ref.~\cite{Epelbaum:2025aan} into the source. The original source is a local Gaussian with $r_0=1.5~{\rm fm}$. Alice denotes the original case, while Bob-I and Bob-II denote the two transformed cases.
}
\label{Fig.InducedNonlocalSource}
\end{figure*}

The scattering-equivalent unitary transformations considered in Ref.~\cite{Epelbaum:2025aan} provide an explicit illustration. They change the interaction-side representation while preserving the on-shell scattering observables. Keeping the original local source fixed while transforming the interaction-side representation produces a source--interaction mismatch. 
Equivalently, with the interaction side kept in its original representation, the mismatch can be absorbed into the source as an induced source-kernel change $\Delta S_U(\mathbf r,\mathbf r')\equiv S_{\rm Bob}(\mathbf r,\mathbf r')-S_{\rm Alice}(\mathbf r,\mathbf r')$ shown in Fig.~\ref{Fig.InducedNonlocalSource}, where $S_{\rm Alice}$ denotes the original source kernel and $S_{\rm Bob}$ a transformed one. We quantify its nonlocal extent by
\begin{equation}\label{Eq.NonlocalKernelWidth}
\sigma_{\mathbf r-\mathbf r'}
=
\left[
\frac{
\displaystyle
\int
\mathrm{d}^3r\mathrm{d}^3r'
\left|\mathbf r-\mathbf r'\right|^2
\left|\Delta S_U(\mathbf r,\mathbf r')\right|^2
}{
\displaystyle
\int
\mathrm{d}^3r\mathrm{d}^3r'
\left|\Delta S_U(\mathbf r,\mathbf r')\right|^2
}
\right]^{1/2}.
\end{equation}
For the two transformed cases, $\sigma_{\mathbf r-\mathbf r'}=2.82$ and $2.53~{\rm fm}$, respectively, both substantially larger than the original Gaussian source radius $r_0=1.5~{\rm fm}$, showing that the induced change is broadly distributed in $\mathbf r-\mathbf r'$.
The resulting difference between the transformed and reference correlation functions reported in Ref.~\cite{Epelbaum:2025aan} shows that an improperly treated mismatch can have a sizable observable effect, rather than signaling an intrinsic shortcoming of femtoscopy.

This raises the practical question of how a consistent source--interaction pairing can be established when no common microscopic derivation is available. We therefore introduce an operational source--interaction convention: a prescription for fixing the source-side representation and establishing a compatible interaction-side representation. Within this convention, the local Gaussian ansatz is imposed in the chosen reference basis. With the source basis fixed, the on-shell scattering observables of a reference pair first constrain the relevant scattering-equivalence class, while fitting its measured correlation function operationally calibrates a compatible interaction-side realization within that class. This establishes an effective source--interaction pairing. 
Since exact microscopic matching cannot generally be established, we regard the pairing as effective if the residual mismatch changes the correlation function by less than the relevant experimental and theoretical accuracies over the kinematic range of interest. Once a theoretical model is calibrated against the measured correlation function of the reference pair, interactions for other particle pairs derived consistently from the same model inherit the same interaction-side convention, enabling CF-to-CF predictions.

Weak sensitivity to extremely short-range structures is not required for the calibration itself, but provides additional robustness against moderate residual mismatch. Realistic off-shell or interaction-model variations have been found to change the correlation function by about $6\%$ or less~\cite{Gobel:2025afq,Molina:2025lzw}. The ALICE analyses further show that the relative contribution of the sub-femtometer region to the source-weighted correlation integral is limited~\cite{ALICE:2023sjd,ALICE:2026tqh}, where wave functions in different interaction-side representations are expected to differ most. Together, these results indicate that moderate residual mismatch need not produce a large net effect on the correlation function in realistic femtoscopic applications.
Returning to the example of Ref.~\cite{Epelbaum:2025aan}, this mismatch lies beyond the validity range of the calibrated effective pairing.

\textit{Conclusions and outlook:} 
Starting from the density operator of the full system, we have formulated two-particle femtoscopic correlations in a density-operator--measurement structure. The source-side object is the reduced density operator of the observed subsystem, while the FSI, or more generally the post-production evolution, is absorbed into an effective measurement operator. The Koonin-Pratt formula is recovered after the sudden, equal-time, on-shell, and diagonal/decoherence reductions, with the conventional source function identified as the diagonal limit of a more general nonlocal density matrix. After the first three reductions, Eq.~\eqref{Eq.GeneralizedRelativeDensityCF} remains well defined without a local classical source interpretation~\cite{Lisa:2005dd} and retains the off-diagonal source structure, or equivalently the relative-momentum dependence of its Wigner representation. This makes the diagonal/decoherence approximation an explicit source-side assumption whose validity can be assessed, particularly in small systems such as $pp$ and $e^+e^-$ collisions.

The operator formulation also makes explicit the need for consistent source--interaction representation matching. In phenomenological applications, this matching is nontrivial because the two sides are constructed independently, whereas on-shell scattering data alone do not uniquely select a compatible interaction-side representation.

The example of Ref.~\cite{Epelbaum:2025aan} shows that a source--interaction mismatch can produce a sizable correlation-level effect rather than an intrinsic shortcoming of femtoscopy. As a practical prescription, we introduced an operational convention that uses on-shell scattering information and the measured correlation function of a reference pair to establish an effective source--interaction pairing for a given theoretical interaction model. Interactions for other particle pairs derived consistently from the same model may then be used for CF-to-CF predictions under the same convention.

Finally, the density-operator--measurement structure provides a natural language for source-side quantum correlations, including coherence and entanglement; extensions to many-body cases; and extensions beyond Koonin-Pratt momentum correlations by choosing appropriate angular or harmonic measurement operators to study angular correlations and flow harmonics~\cite{Voloshin:1994mz, Poskanzer:1998yz,Bilandzic:2013kga}.

%=========================================================================&
%=========================================================================&
\textit{Acknowledgments: }
This work is partly supported by the National Natural Science Foundation of China under Grants No.W2543006 and No.12435007.
Zhi-Wei Liu acknowledges support from the National Natural Science Foundation of China under Grant No.12405133, and the Shenzhen Science and Technology Program under Grant No.ZDSYS20230626091501002. H.N.L. thanks 
the China Scholarship Council for supporting his visit to USP.

%=========================================================================&
%=========================================================================&
%=========================================================================&

%=========================================================================&
%=========================================================================&
%=========================================================================&

\nocite{Fano:1957zz,nielsen2010quantum,von2013mathematische}
%,Smith:2026ocz,Lisa:2005dd,Pratt:1990zq,Lednicky:2007ax,Pratt:1997pw

\bibliographystyle{apsrev4-1}
\bibliography{rho2CF}

@article{Pratt:1990zq,
    author = "Pratt, S. and Csorgo, T. and Zimanyi, J.",
    title = "{Detailed predictions for two pion correlations in ultrarelativistic heavy ion collisions}",
    doi = "10.1103/PhysRevC.42.2646",
    journal = "Phys. Rev. C",
    volume = "42",
    pages = "2646--2652",
    year = "1990"
}

@article{Smith:2026ocz,
    author = "Smith, Isaac G. and Blum, Kfir",
    title = "{Corrections to the smoothness and on-shell approximations in femtoscopy and coalescence}",
    eprint = "2602.02810",
    archivePrefix = "arXiv",
    primaryClass = "nucl-th",
    doi = "10.1103/z5pg-7hwx",
    journal = "Phys. Rev. C",
    volume = "114",
    number = "1",
    pages = "014908",
    year = "2026"
}

@article{Lednicky:2005tb,
    author = "Lednicky, Richard",
    title = "{Finite-size effects on two-particle production in continuous and discrete spectrum}",
    eprint = "nucl-th/0501065",
    archivePrefix = "arXiv",
    reportNumber = "DIRAC-2004-06, DIRAC Note 2004-06, CERN, 27.11.2004",
    doi = "10.1134/S1063779609030034",
    journal = "Phys. Part. Nucl.",
    volume = "40",
    pages = "307--352",
    year = "2009"
}

@article{Epelbaum:2025aan,
    author = "Epelbaum, Evgeny and Heihoff, Sven and Mei{\ss}ner, Ulf-G. and Tscherwon, Alexander",
    title = "{Can the Strong Interactions between Hadrons Be Determined Using Femtoscopy?}",
    eprint = "2504.08631",
    archivePrefix = "arXiv",
    primaryClass = "nucl-th",
    doi = "10.1103/tfsb-wlsd",
    journal = "Phys. Rev. Lett.",
    volume = "136",
    number = "21",
    pages = "212301",
    year = "2026"
}

@article{kraus1971general,
  title={General state changes in quantum theory},
  author={Kraus, Karl},
  journal={Annals of Physics},
  volume={64},
  number={2},
  pages={311--335},
  year={1971},
  publisher={Elsevier}
}

@book{kraus1983states,
  title={States, effects, and operations fundamental notions of quantum theory: Lectures in mathematical physics at the university of Texas at Austin},
  author={Kraus, Karl and B{\"o}hm, Arno and Dollard, John D and Wootters, WH},
  year={1983},
  publisher={Springer}
}

@article{choi1975completely,
  title={Completely positive linear maps on complex matrices},
  author={Choi, Man-Duen},
  journal={Linear algebra and its applications},
  volume={10},
  number={3},
  pages={285--290},
  year={1975},
  publisher={Elsevier}
}

@article{Stinespring:1955eig,
    author = "Stinespring, W. Forrest",
    title = "{Positive functions on {\ensuremath{\mathit{C}}}*-algebras}",
    doi = "10.1090/s0002-9939-1955-0069403-4",
    journal = "Proc. Am. Math. Soc.",
    volume = "6",
    number = "2",
    pages = "211--216",
    year = "1955"
}

@article{Hirano:2005xf,
    author = "Hirano, Tetsufumi and Heinz, Ulrich W. and Kharzeev, Dmitri and Lacey, Roy and Nara, Yasushi",
    title = "{Hadronic dissipative effects on elliptic flow in ultrarelativistic heavy-ion collisions}",
    eprint = "nucl-th/0511046",
    archivePrefix = "arXiv",
    doi = "10.1016/j.physletb.2006.03.060",
    journal = "Phys. Lett. B",
    volume = "636",
    pages = "299--304",
    year = "2006"
}

@article{Lin:2002gc,
    author = "Lin, Zi-wei and Ko, C. M. and Pal, Subrata",
    title = "{Partonic effects on pion interferometry at RHIC}",
    eprint = "nucl-th/0204054",
    archivePrefix = "arXiv",
    doi = "10.1103/PhysRevLett.89.152301",
    journal = "Phys. Rev. Lett.",
    volume = "89",
    pages = "152301",
    year = "2002"
}

@book{breuer2002theory,
  title={The theory of open quantum systems},
  author={Breuer, Heinz-Peter and Petruccione, Francesco},
  year={2002},
  publisher={OUP Oxford}
}

@article{Koonin:1977fh,
    author = "Koonin, S. E.",
    title = "{Proton Pictures of High-Energy Nuclear Collisions}",
    doi = "10.1016/0370-2693(77)90340-9",
    journal = "Phys. Lett. B",
    volume = "70",
    pages = "43--47",
    year = "1977"
}

@article{Lisa:2005dd,
    author = "Lisa, Michael Annan and Pratt, Scott and Soltz, Ron and Wiedemann, Urs",
    title = "{Femtoscopy in relativistic heavy ion collisions}",
    eprint = "nucl-ex/0505014",
    archivePrefix = "arXiv",
    doi = "10.1146/annurev.nucl.55.090704.151533",
    journal = "Ann. Rev. Nucl. Part. Sci.",
    volume = "55",
    pages = "357--402",
    year = "2005"
}

@book{rivas2012open,
  title={Open quantum systems},
  author={Rivas, Angel and Huelga, Susana F},
  volume={10},
  year={2012},
  publisher={Springer}
}

@article{Wiedemann:1999qn,
    author = "Wiedemann, Urs Achim and Heinz, Ulrich W.",
    title = "{Particle interferometry for relativistic heavy ion collisions}",
    eprint = "nucl-th/9901094",
    archivePrefix = "arXiv",
    reportNumber = "CU-TP-931, CERN-TH-99-15",
    doi = "10.1016/S0370-1573(99)00032-0",
    journal = "Phys. Rept.",
    volume = "319",
    pages = "145--230",
    year = "1999"
}

@article{Boal:1990yh,
    author = "Boal, D. H. and Gelbke, C. K. and Jennings, B. K.",
    title = "{Intensity interferometry in subatomic physics}",
    doi = "10.1103/RevModPhys.62.553",
    journal = "Rev. Mod. Phys.",
    volume = "62",
    pages = "553--602",
    year = "1990"
}

@article{Weiner:1999th,
    author = "Weiner, R. M.",
    title = "{Boson interferometry in high-energy physics}",
    eprint = "hep-ph/9904389",
    archivePrefix = "arXiv",
    doi = "10.1016/S0370-1573(99)00114-3",
    journal = "Phys. Rept.",
    volume = "327",
    pages = "249--346",
    year = "2000"
}

@article{Fabbietti:2020bfg,
    author = "Fabbietti, L. and Mantovani Sarti, V. and Vazquez Doce, O.",
    title = "{Study of the Strong Interaction Among Hadrons with Correlations at the LHC}",
    eprint = "2012.09806",
    archivePrefix = "arXiv",
    primaryClass = "nucl-ex",
    doi = "10.1146/annurev-nucl-102419-034438",
    journal = "Ann. Rev. Nucl. Part. Sci.",
    volume = "71",
    pages = "377--402",
    year = "2021"
}

@article{ALICE:2018ysd,
    author = "Acharya, Shreyasi and others",
    collaboration = "ALICE",
    title = "{p-p, p-$\Lambda$ and $\Lambda$-$\Lambda$ correlations studied via femtoscopy in pp reactions at $\sqrt{s}$ = 7 TeV}",
    eprint = "1805.12455",
    archivePrefix = "arXiv",
    primaryClass = "nucl-ex",
    reportNumber = "CERN-EP-2018-150",
    doi = "10.1103/PhysRevC.99.024001",
    journal = "Phys. Rev. C",
    volume = "99",
    number = "2",
    pages = "024001",
    year = "2019"
}

@article{ALICE:2019hdt,
    author = "Acharya, Shreyasi and others",
    collaboration = "ALICE",
    title = "{First Observation of an Attractive Interaction between a Proton and a Cascade Baryon}",
    eprint = "1904.12198",
    archivePrefix = "arXiv",
    primaryClass = "nucl-ex",
    reportNumber = "CERN-EP-2019-080",
    doi = "10.1103/PhysRevLett.123.112002",
    journal = "Phys. Rev. Lett.",
    volume = "123",
    number = "11",
    pages = "112002",
    year = "2019"
}

@article{ALICE:2020mfd,
    author = "Collaboration, Alice and others",
    collaboration = "ALICE",
    title = "{Unveiling the strong interaction among hadrons at the LHC}",
    eprint = "2005.11495",
    archivePrefix = "arXiv",
    primaryClass = "nucl-ex",
    reportNumber = "CERN-EP-2020-091",
    doi = "10.1038/s41586-020-3001-6",
    journal = "Nature",
    volume = "588",
    pages = "232--238",
    year = "2020",
    note = "[Erratum: Nature 590, E13 (2021)]"
}

@article{STAR:2015kha,
    author = "Adamczyk, L. and others",
    collaboration = "STAR",
    title = "{Measurement of Interaction between Antiprotons}",
    eprint = "1507.07158",
    archivePrefix = "arXiv",
    primaryClass = "nucl-ex",
    doi = "10.1038/nature15724",
    journal = "Nature",
    volume = "527",
    pages = "345--348",
    year = "2015"
}

@article{Si:2025eou,
    author = "Si, Dawei and others",
    title = "{Extracting Neutron-Neutron Interaction Strength and Spatiotemporal Dynamics of Neutron Emission from the Two-Particle Correlation Function}",
    eprint = "2501.09576",
    archivePrefix = "arXiv",
    primaryClass = "nucl-ex",
    doi = "10.1103/PhysRevLett.134.222301",
    journal = "Phys. Rev. Lett.",
    volume = "134",
    number = "22",
    pages = "222301",
    year = "2025"
}

@article{Fano:1957zz,
    author = "Fano, U.",
    title = "{Description of States in Quantum Mechanics by Density Matrix and Operator Techniques}",
    doi = "10.1103/RevModPhys.29.74",
    journal = "Rev. Mod. Phys.",
    volume = "29",
    pages = "74--93",
    year = "1957"
}

@book{nielsen2010quantum,
  title={Quantum computation and quantum information},
  author={Nielsen, Michael A and Chuang, Isaac L},
  year={2010},
  publisher={Cambridge university press}
}

@book{von2013mathematische,
  title={Mathematische grundlagen der quantenmechanik},
  author={Von Neumann, John},
  volume={38},
  year={2013},
  publisher={Springer-Verlag}
}

@article{Chapman:1994xa,
    author = "Chapman, Scott and Heinz, Ulrich W.",
    title = "{HBT correlators: Current formalism versus Wigner function formulation}",
    eprint = "hep-ph/9407405",
    archivePrefix = "arXiv",
    doi = "10.1016/0370-2693(94)01277-6",
    journal = "Phys. Lett. B",
    volume = "340",
    pages = "250--253",
    year = "1994"
}

@mastersthesis{Mihaylov:2021glj,
    author = "Mihaylov, Dimitar Lubomirov",
    title = "{Analysis techniques for femtoscopy and correlation studies in small collision systems and their applications to the investigation of p{\textendash}{\ensuremath{\Lambda}} and {\ensuremath{\Lambda}}{\textendash}{\ensuremath{\Lambda}} interactions with ALICE}",
    reportNumber = "CERN-THESIS-2021-052",
    type = "Other thesis",
    school = "2021-02-16, Munich, Tech. U.",
    month = "2",
    year = "2021"
}

@article{Bialas:2005ps,
    author = "Bialas, A. and Zalewski, K.",
    title = "{Bose-Einstein correlations: A Study of an invariance group}",
    eprint = "hep-ph/0501017",
    archivePrefix = "arXiv",
    doi = "10.1103/PhysRevD.72.036009",
    journal = "Phys. Rev. D",
    volume = "72",
    pages = "036009",
    year = "2005"
}

@article{Pratt:1997pw,
    author = "Pratt, S.",
    title = "{Validity of the smoothness assumption for calculating two-boson correlations in high-energy collisions}",
    doi = "10.1103/PhysRevC.56.1095",
    journal = "Phys. Rev. C",
    volume = "56",
    pages = "1095--1098",
    year = "1997"
}

@article{Bauer:1992ffu,
    author = "Bauer, W. and Gelbke, C. K. and Pratt, S.",
    title = "{Hadronic interferometry in heavy ion collisions}",
    doi = "10.1146/annurev.ns.42.120192.000453",
    journal = "Ann. Rev. Nucl. Part. Sci.",
    volume = "42",
    pages = "77--100",
    year = "1992"
}

@article{Csorgo:1995bi,
    author = "Csorgo, T. and Lorstad, B.",
    title = "{Bose-Einstein correlations for three-dimensionally expanding, cylindrically symmetric, finite systems}",
    eprint = "hep-ph/9509213",
    archivePrefix = "arXiv",
    reportNumber = "LUNFD6-NFFL-7082-REV, LUNFD6-NFFL-7082",
    doi = "10.1103/PhysRevC.54.1390",
    journal = "Phys. Rev. C",
    volume = "54",
    pages = "1390--1403",
    year = "1996"
}

@article{ALICE:2011dyt,
    author = "Aamodt, K. and others",
    collaboration = "ALICE",
    title = "{Two-pion Bose-Einstein correlations in central Pb-Pb collisions at $\sqrt{{s}_{NN}} =$ 2.76 TeV}",
    eprint = "1012.4035",
    archivePrefix = "arXiv",
    primaryClass = "nucl-ex",
    reportNumber = "CERN-PH-EP-ALICE-2010-006, CERN-PH-EP-2010-078",
    doi = "10.1016/j.physletb.2010.12.053",
    journal = "Phys. Lett. B",
    volume = "696",
    pages = "328--337",
    year = "2011"
}

@article{STAR:2004qya,
    author = "Adams, J. and others",
    collaboration = "STAR",
    title = "{Pion interferometry in Au+Au collisions at S(NN)**(1/2) = 200-GeV}",
    eprint = "nucl-ex/0411036",
    archivePrefix = "arXiv",
    doi = "10.1103/PhysRevC.71.044906",
    journal = "Phys. Rev. C",
    volume = "71",
    pages = "044906",
    year = "2005"
}

@article{STAR:2009fks,
    author = "Abelev, B. I. and others",
    collaboration = "STAR",
    title = "{Pion Interferometry in Au+Au and Cu+Cu Collisions at RHIC}",
    eprint = "0903.1296",
    archivePrefix = "arXiv",
    primaryClass = "nucl-ex",
    doi = "10.1103/PhysRevC.80.024905",
    journal = "Phys. Rev. C",
    volume = "80",
    pages = "024905",
    year = "2009"
}

@article{HanburyBrown:1956bqd,
    author = "Hanbury Brown, R. and Twiss, R. Q.",
    title = "{A Test of a new type of stellar interferometer on Sirius}",
    doi = "10.1038/1781046a0",
    journal = "Nature",
    volume = "178",
    pages = "1046--1048",
    year = "1956"
}

@article{Voloshin:1994mz,
    author = "Voloshin, S. and Zhang, Y.",
    title = "{Flow study in relativistic nuclear collisions by Fourier expansion of Azimuthal particle distributions}",
    eprint = "hep-ph/9407282",
    archivePrefix = "arXiv",
    doi = "10.1007/s002880050141",
    journal = "Z. Phys. C",
    volume = "70",
    pages = "665--672",
    year = "1996"
}

@article{Poskanzer:1998yz,
    author = "Poskanzer, Arthur M. and Voloshin, S. A.",
    title = "{Methods for analyzing anisotropic flow in relativistic nuclear collisions}",
    eprint = "nucl-ex/9805001",
    archivePrefix = "arXiv",
    doi = "10.1103/PhysRevC.58.1671",
    journal = "Phys. Rev. C",
    volume = "58",
    pages = "1671--1678",
    year = "1998"
}

@article{Bilandzic:2013kga,
    author = "Bilandzic, Ante and Christensen, Christian Holm and Gulbrandsen, Kristjan and Hansen, Alexander and Zhou, You",
    title = "{Generic framework for anisotropic flow analyses with multiparticle azimuthal correlations}",
    eprint = "1312.3572",
    archivePrefix = "arXiv",
    primaryClass = "nucl-ex",
    doi = "10.1103/PhysRevC.89.064904",
    journal = "Phys. Rev. C",
    volume = "89",
    number = "6",
    pages = "064904",
    year = "2014"
}

@article{Lednicky:2007ax,
    author = "Lednicky, Richard",
    editor = "Padula, Sandra S. and Pratt, Scott",
    title = "{Femtoscopic correlations in multiparticle production and beta-decay}",
    eprint = "nucl-th/0702063",
    archivePrefix = "arXiv",
    doi = "10.1590/S0103-97332007000600011",
    journal = "Braz. J. Phys.",
    volume = "37",
    pages = "939--948",
    year = "2007"
}

@article{ALICE:2023sjd,
    author = "Acharya, Shreyasi and others",
    collaboration = "ALICE",
    title = "{Common femtoscopic hadron-emission source in pp collisions at the LHC}",
    eprint = "2311.14527",
    archivePrefix = "arXiv",
    primaryClass = "hep-ph",
    reportNumber = "CERN-EP-2023-267",
    doi = "10.1140/epjc/s10052-025-13793-y",
    journal = "Eur. Phys. J. C",
    volume = "85",
    number = "2",
    pages = "198",
    year = "2025",
    note = "[Erratum: Eur.Phys.J.C 86, 12 (2026)]"
}

@article{ALICE:2026tqh,
    author = "Ali Hassan Abdallah, Dana and others",
    collaboration = "ALICE",
    title = "{Multiplicity dependence of the size of the common hadron emission source in pp collisions at the LHC}",
    eprint = "2606.28098",
    archivePrefix = "arXiv",
    primaryClass = "nucl-ex",
    reportNumber = "CERN-EP-2026-180",
    month = "6",
    year = "2026"
}

@article{Molina:2025lzw,
    author = "Molina, R. and Oset, E.",
    title = "{Determination of off-shell ambiguities in correlation functions: Strategies to minimize them}",
    eprint = "2506.03669",
    archivePrefix = "arXiv",
    primaryClass = "hep-ph",
    doi = "10.1103/rst4-rkmm",
    journal = "Phys. Rev. D",
    volume = "112",
    number = "9",
    pages = "096006",
    year = "2025"
}

@article{Gobel:2025afq,
    author = {G{\"o}bel, Matthias and Kievsky, Alejandro},
    title = "{Nucleon-nucleon correlation functions from different interactions in comparison}",
    eprint = "2505.13433",
    archivePrefix = "arXiv",
    primaryClass = "nucl-th",
    doi = "10.1016/j.physletb.2025.139835",
    journal = "Phys. Lett. B",
    volume = "869",
    pages = "139835",
    year = "2025"
}

@article{Coci:2026puh,
    author = "Coci, Gabriele and Plumari, Salvatore and Falci, Giuseppe",
    title = "{Quantum decoherence: a study applied to quarkonium-like bound states in strongly interacting matter}",
    eprint = "2607.06137",
    archivePrefix = "arXiv",
    primaryClass = "hep-ph",
    month = "7",
    year = "2026"
}

@misc{SupplementalMaterial,
  note = {See Supplemental Material at [URL will be inserted by publisher] for details on the derivation of the formalism, the relation between the diagonal and smoothness approximations, source--interaction matching, the operational convention, and the coupled-channel extension, which includes Refs. [44--46].}
}
%=========================================================================&
%=========================================================================&
%=========================================================================&

\clearpage
\onecolumngrid

% Supplemental Material
\setcounter{secnumdepth}{0}
\setcounter{equation}{0}
\renewcommand{\theequation}{SM.\arabic{equation}}

\begin{center}
{\large\bfseries Supplemental Material}
\end{center}

In this Supplemental Material, we present some details regarding the recovery of the primitive traditional formulation of correlation functions, the relation between the diagonal and smoothness approximations, the representation-matching requirement between the source and interaction sectors, the induced nonlocal source kernel associated with an uncontrolled mismatch, the operational source--interaction convention, and the extension to coupled-channel quantum maps and channel-projected correlation functions.

\vspace{1em}

\section{Derivation of the primitive representation of the traditional framework from our framework}\label{Sec:PrimitiveEquivalence}
This section makes explicit the connection between the density-operator formulation developed in the main text and the primitive amplitude-level formulation of the traditional framework. We show that the latter can be recovered from the former once the reduced two-body density operator is represented as an ensemble of source amplitudes, with the usual product form of the source appearing as an additional factorization assumption.

We begin from the primitive starting point of the traditional expression, namely the two-particle probability introduced by Pratt, Cs\"org\H{o}, and Zim\'anyi~\cite{Pratt:1990zq}. In that formulation, the two-particle spectrum can be written as
\begin{equation}\label{Eq.Pratt}
P(p_1,p_2)=
\sum_F
\left|
\int \mathrm{d}^4x_1\,\mathrm{d}^4x_2\;
T_{F_1}(x_1)\,T_{F_2}(x_2)\,
U(x_1,x_2,p_1,p_2;t=\infty)
\right|^2,
\end{equation}
where $T_{F_i}(x_i)$ denotes the $T$ matrix between the observed particle $i$ and the residual system (labeled by $F_i$) at the space-time point $x_i$, while $U(x_1,x_2,p_1,p_2;t=\infty)$ represents the propagation from the emission points to the detector.

As we will see, the above expression emerges naturally from the present framework. In high-energy collisions, the full system may be idealized as a pure quantum state before any coarse-graining or tracing over unobserved degrees of freedom is performed. It can therefore be represented by a state vector. However, the observed $n$-particle subsystem is typically mixed and must be treated with a density matrix~\cite{von2013mathematische,Fano:1957zz,nielsen2010quantum}. For a composite system consisting of the observed $n$-body subsystem and its environment, one can always perform a Schmidt decomposition of the overall state vector on $\mathcal H_n\otimes\mathcal H_E$
\begin{equation}
\ket{\Psi}
=
\sum_F \sqrt{P_F}\,\ket{\psi_F}\otimes\ket{\phi_F},
\end{equation}
where $\ket{\phi_F}$ denotes the environmental state, while $\ket{\psi_F}$ denotes the corresponding state of the observed $n$-body subsystem. The coefficients $P_F$ are non-negative and normalized as
\begin{equation}
\sum_F P_F=1.
\end{equation}

Specializing now to the two-body case, tracing over the environmental degrees of freedom yields
\begin{equation}
\hat\rho^{(2)}_{\mathrm{in}}
=
\mathrm{Tr}_{\mathcal H_E}\ket{\Psi}\bra{\Psi}
=
\sum_F P_F\,\ket{\psi_F}\bra{\psi_F}.
\end{equation}

Starting from Eq.~\eqref{Eq.ParentFormula_with_SuddenApproximation} of the main text, and inserting complete sets of two-body coordinate states, one obtains
\begin{align}\label{Eq.MyFrame2TradFrame}
C(p_1,p_2)
&=
\sum_F P_F
\int \mathrm{d}^4x_1\,\mathrm{d}^4x_2\,\mathrm{d}^4x_1'\,\mathrm{d}^4x_2'\;
\bra{p_1,p_2}\hat U_{\rm FSI}^{(2)}\ket{x_1,x_2}
\braket{x_1,x_2|\psi_F}
\braket{\psi_F|x_1',x_2'}
\bra{x_1',x_2'}\hat U_{\rm FSI}^{(2)\dagger}\ket{p_1,p_2}.
\end{align}

Defining $T_F(x_1,x_2)=\sqrt{P_F}\,\braket{x_1,x_2|\psi_F}$ and $U(x_1,x_2,p_1,p_2)=\bra{p_1,p_2}\hat U_{\rm FSI}^{(2)}\ket{x_1,x_2}$, Eq.~\eqref{Eq.MyFrame2TradFrame} becomes
\begin{equation}
C(p_1,p_2)
=
\sum_F
\left|
\int \mathrm{d}^4x_1\,\mathrm{d}^4x_2\;
T_F(x_1,x_2)\,
U(x_1,x_2,p_1,p_2)
\right|^2.
\end{equation}

It should be emphasized that, although the notation is introduced here within our density operator framework, the functions $T_F$ and $U$ defined above play the same roles as their counterparts in Eq.~\eqref{Eq.Pratt}. The structure is therefore identical to the traditional starting point, except that here the source-side object is the more general two-body object $T_F(x_1,x_2)$ rather than the factorized product $T_{F_1}(x_1)T_{F_2}(x_2)$. If one imposes the additional assumption that the two-body object can be factorized into a product of one-body objects,
\begin{equation}
T_F(x_1,x_2)\simeq T_{F_1}(x_1)\,T_{F_2}(x_2),
\end{equation}
the present formula immediately reduces to the traditional one. Thus, the traditional form is recovered after factorization, while the present framework retains the more general two-body structure at the outset.

\section{Relation between the diagonal approximation and the smoothness approximation}
The diagonal approximation introduced in the main text has a direct counterpart in the conventional Wigner-source formulation. To see this connection, we perform the Wigner transformation of the equal-time density matrix~\cite{Lednicky:2007ax}
\begin{equation}\label{Eq.WignerTransformation}
S_{\mathbf P}(\mathbf R,\mathbf k)
=
\int \mathrm{d}^3s\,
e^{-i\mathbf k\cdot\mathbf s}\,
\rho_{\mathbf P}
\left(
\mathbf R+\frac{\mathbf s}{2},
\mathbf R-\frac{\mathbf s}{2}
\right).
\end{equation}
This transformation converts the nonlocal source, written in terms of two relative coordinates, into a phase-space source depending on both the relative position $\mathbf R=(\mathbf r +\mathbf r')/2$ and the relative momentum $\mathbf k$. Therefore, the off-diagonal dependence of $\rho_{\mathbf P}(\mathbf r,\mathbf r')$ on $\mathbf s=\mathbf r-\mathbf r'$ is equivalently encoded in the $\mathbf k$-dependence of the Wigner source $S_{\mathbf P}(\mathbf R,\mathbf k)$. The smoothness approximation removes the $\mathbf k$-dependence of $S_{\mathbf P}(\mathbf R,\mathbf k)$, and Eq.~\eqref{Eq.WignerTransformation} then reduces to Eq.~\eqref{Eq.DiagonalApproximation} of the main text.
The smoothness approximation and the diagonal approximation should therefore be regarded as two closely related but representation-dependent source-side reductions. In the Wigner representation, the reduction appears as the suppression of the relative-momentum dependence of the emission function~\cite{Smith:2026ocz}. In the relative-coordinate density-matrix representation, the corresponding loss of resolvable source-side interference is represented by the suppression of off-diagonal matrix elements. This distinction is useful because the generalized correlation function in Eq.~\eqref{Eq.GeneralizedRelativeDensityCF} is already well defined before the diagonal approximation is introduced. It therefore does not require the smoothness approximation to be used as a correlation formula. Conversely, the inverse Wigner transformation makes explicit the relationship between relative-momentum dependence and the off-diagonal structure of the relative-coordinate density matrix. With the present convention, one obtains
\begin{equation}
\rho_{\mathbf P}
\left(
\mathbf R+\frac{\mathbf s}{2},
\mathbf R-\frac{\mathbf s}{2}
\right)=
\int \frac{\mathrm{d}^3k}{(2\pi)^3}
e^{i\mathbf k\cdot\mathbf s}
S_{\mathbf P}(\mathbf R,\mathbf k).
\end{equation}

Thus the two-particle phase-space distribution obtained from transport models can, at least at the semiclassical level, provide information about the off-diagonal structure of the relative-coordinate density matrix.

More quantitatively, the quality of the diagonal approximation can be characterized by the effective off-diagonal width $\Delta r$ of the density matrix in the relative-coordinate difference $\mathbf s=\mathbf r-\mathbf r'$. In Eq.~\eqref{Eq.WignerTransformation} we have set $\hbar=1$. Restoring $\hbar$, or equivalently working in SI units with $\mathbf k$ interpreted as the physical relative momentum, this width is controlled by the momentum scale over which the Wigner source varies,
\begin{equation}
\Delta r \sim \frac{\hbar}{\Delta k}.
\end{equation}

There are therefore two formal ways to drive the density matrix toward a diagonal form. One is the classical limit $\hbar\to0$ at fixed physical momentum scale $\Delta k$. The other is the extreme smoothness limit $\Delta k\to\infty$ at fixed $\hbar$, in which the Wigner source becomes effectively flat in the relative momentum over the relevant region. In practice, however, neither limit has to be taken literally. It is enough to have the scale separation
\begin{equation}
\Delta r \ll L_{\rm res},
\qquad \text{or equivalently} \qquad
\Delta k \gg \frac{\hbar}{L_{\rm res}},
\end{equation}
where $L_{\rm res}$ denotes the characteristic $\mathbf s$-variation scale of the relative wave function, or more generally of the measurement kernel entering the correlation function. The first form states that the off-diagonal width of the density matrix is too small to be resolved by the correlation measurement, while the second expresses the same condition in the Wigner representation: the source must vary only over a momentum scale much broader than the relative-momentum scale probed by the pair. In this sense, the smoothness approximation realizes the diagonal approximation by taking the effective momentum width $\Delta k$ to be large, whereas the classical limit realizes the same reduction by taking $\hbar$ to be small.

This is particularly relevant for small systems, such as $pp$ or $e^+e^-$ collisions, where the relative-momentum dependence of the source may be less safely neglected~\cite{Pratt:1997pw,Lisa:2005dd}.

\section{Source--interaction matching}
\label{App:Mismatch}

In the density-operator formulation, the two-particle correlation function is
\begin{equation}\label{Eq.operatorlevel}
C(\mathbf q,\mathbf P)
=
\langle\psi_{\mathbf q}^{(-)}|
\hat\rho_{\mathbf P}
|\psi_{\mathbf q}^{(-)}\rangle .
\end{equation}
Inserting two complete sets of continuous relative-coordinate states gives
\begin{equation}\label{Eq.representationlevel}
C(\mathbf q,\mathbf P)
=
\int \mathrm d^3r\,\mathrm d^3r'\,
\langle\psi_{\mathbf q}^{(-)}|\mathbf r\rangle
\langle\mathbf r|\hat\rho_{\mathbf P}|\mathbf r'\rangle
\langle\mathbf r'|\psi_{\mathbf q}^{(-)}\rangle .
\end{equation}

At the operator level, Eq.~\eqref{Eq.operatorlevel} is independent of any particular basis. In practice, however, evaluating Eq.~\eqref{Eq.representationlevel} requires choosing a basis in which the density operator and scattering state are represented. In particular, diagonality is a property of the matrix representation of a density operator in a specified basis, rather than of the abstract operator itself.

To make this distinction explicit, consider the simultaneous transformation of the operator and state, together with an optional corresponding transformation of the coordinate basis,
\begin{equation}
\widetilde{\hat\rho}
=
\hat U\hat\rho\hat U^\dagger,
\qquad
|\widetilde\psi\rangle
=
\hat U|\psi\rangle,
\qquad
|\widetilde{\mathbf r}\rangle
=
\hat U|\mathbf r\rangle .
\end{equation}
The first two relations define the operator-level transformation, whereas the third is an optional accompanying transformation of the coordinate basis.

If the operator and state are transformed together with the basis, their coordinate-space matrix elements remain unchanged:
\begin{equation}
\langle\widetilde{\mathbf r}|
\widetilde{\hat\rho}
|\widetilde{\mathbf r}'\rangle
=
\langle\mathbf r|
\hat\rho
|\mathbf r'\rangle,
\qquad
\langle\widetilde{\mathbf r}|\widetilde\psi\rangle
=
\langle\mathbf r|\psi\rangle .
\end{equation}
Independently of whether the basis is transformed, the operator-level expectation value is invariant,
\begin{equation}
\langle\widetilde\psi|
\widetilde{\hat\rho}
|\widetilde\psi\rangle
=
\langle\psi|
\hat\rho
|\psi\rangle .
\end{equation}
Thus, the simultaneous unitary transformation of the density operator and state is sufficient to leave the expectation value invariant. Transforming the basis as well keeps their coordinate-space representations unchanged.

If, instead, the transformed operator and state are represented in the original fixed basis $\{|\mathbf r\rangle\}$, their matrix elements generally change:
\begin{align}
\langle\psi|\hat\rho|\psi\rangle
&=
\langle\psi|
\hat U^\dagger
\bigl(\hat U\hat\rho\hat U^\dagger\bigr)
\hat U
|\psi\rangle
\nonumber\\
&=
\int \mathrm d^3r\,\mathrm d^3r'\,
\langle\psi|\hat U^\dagger|\mathbf r\rangle
\langle\mathbf r|
\hat U\hat\rho\hat U^\dagger
|\mathbf r'\rangle
\langle\mathbf r'|\hat U|\psi\rangle .
\end{align}
Their matrix elements then become
\begin{equation}
\langle\mathbf r|
\widetilde{\hat\rho}
|\mathbf r'\rangle
=
\langle\mathbf r|
\hat U\hat\rho\hat U^\dagger
|\mathbf r'\rangle,
\qquad
\langle\mathbf r|\widetilde\psi\rangle
=
\langle\mathbf r|
\hat U|\psi\rangle,
\end{equation}
and generally differ from their original forms. In particular, a source that is diagonal in the original basis need not remain diagonal thereafter.

Equivalently, the transformed wave function $\langle\mathbf r|\widetilde\psi\rangle$ in the original basis may be viewed as the original abstract scattering state represented in the basis
\begin{equation}
|\mathbf r\rangle_U
=
\hat U^\dagger|\mathbf r\rangle .
\end{equation}

A scattering-equivalent unitary transformation preserves the asymptotic scattering content and hence the on-shell observables, while generally changing the interior wave function and other off-shell structures. It may therefore be understood as representing the same physical scattering state in a different basis.

If the source density operator and the scattering state are derived consistently as operators within the same microscopic framework, their matching is implemented automatically at the operator level. Expressing both objects in any common complete basis changes only their matrix representations and leaves their operator-level matching intact. In phenomenological applications, however, the source kernel and scattering wave function are usually constructed independently, so the matching of their representations is not automatically
guaranteed.

To distinguish the basis choices associated with the source and interaction sectors, we denote by
\begin{equation}
\mathcal B_S
=
\left\{
|\mathbf r\rangle_S
\right\}
\end{equation}
the source-side basis in which the density-matrix kernel is modeled,
\begin{equation}
\rho_S(\mathbf r,\mathbf r')
=
{}_S\langle\mathbf r|
\hat\rho
|\mathbf r'\rangle_S,
\qquad
\psi_S(\mathbf r)
=
{}_S\langle\mathbf r|\psi\rangle .
\end{equation}
Likewise, we denote by
\begin{equation}
\mathcal B_I
=
\left\{
|\mathbf r\rangle_I
\right\}
\end{equation}
the coordinate basis associated with the chosen interaction-side representation, in which
\begin{equation}
\rho_I(\mathbf r,\mathbf r')
=
{}_I\langle\mathbf r|
\hat\rho
|\mathbf r'\rangle_I,
\qquad
\psi_I(\mathbf r)
=
{}_I\langle\mathbf r|\psi\rangle .
\end{equation}
Although the same continuous coordinate label $\mathbf r$ is used in both representations, this does not by itself imply that $|\mathbf r\rangle_S$ and $|\mathbf r\rangle_I$ denote the same basis vectors.

If the two bases are related by
\begin{equation}
|\mathbf r\rangle_I
=
\hat U^\dagger|\mathbf r\rangle_S,
\end{equation}
then
\begin{equation}
\psi_I(\mathbf r)
=
{}_S\langle\mathbf r|
\hat U|\psi\rangle,
\qquad
\rho_I(\mathbf r,\mathbf r')
=
{}_S\langle\mathbf r|
\hat U\hat\rho\hat U^\dagger
|\mathbf r'\rangle_S .
\end{equation}
The correlation function can be evaluated consistently in either representation,
\begin{align}
C
&=
\int \mathrm d^3r\,\mathrm d^3r'\,
\psi_S^*(\mathbf r)\,
\rho_S(\mathbf r,\mathbf r')\,
\psi_S(\mathbf r')
\nonumber\\
&=
\int \mathrm d^3r\,\mathrm d^3r'\,
\psi_I^*(\mathbf r)\,
\rho_I(\mathbf r,\mathbf r')\,
\psi_I(\mathbf r') .
\end{align}
By contrast, combining $\rho_S(\mathbf r,\mathbf r')$ directly with $\psi_I(\mathbf r)$ amounts to identifying the two representations without the required transformation and therefore generates a source--interaction mismatch.

This issue is not removed by constraining the interaction with scattering data alone. On-shell scattering observables constrain the on-shell scattering amplitude, but do not in general uniquely fix an off-shell continuation or interior representation. Scattering-equivalent interaction models may therefore be indistinguishable in scattering experiments while corresponding to different interaction-side representations in the correlation calculation. Consequently, an interaction that reproduces the available on-shell scattering data is not, by that fact alone, automatically matched to an independently specified source
representation.

The scattering-equivalent transformations considered in Ref.~\cite{Epelbaum:2025aan} provide an explicit illustration of this issue. In the next section, we represent the resulting mismatch entirely on the source side and quantify the induced nonlocal source kernel.

\section{Off-shell ambiguity and induced nonlocal source kernels}
\label{App:Meissner_response}

We now consider the scattering-equivalent transformations of Ref.~\cite{Epelbaum:2025aan}, which change the interaction-side representation while preserving the on-shell scattering observables. For this comparison, the pre-transformation source--interaction combination serves only as the reference configuration. Relative to this reference, keeping the same local source fixed while transforming the interaction-side representation produces a source--interaction mismatch. Equivalently, this mismatch can be represented entirely in the original interaction-side representation by transforming the source operator. Let $\hat S_A$ be the original source operator, whose coordinate-space kernel is
\begin{equation}
\langle \mathbf r|\hat S_A|\mathbf r'\rangle
=
S_0(r)\delta^{(3)}(\mathbf r-\mathbf r') .
\label{Eq.SM.LocalSourceAlice}
\end{equation}

If the wave function is transformed while the source is kept fixed, one obtains
\begin{equation}
\bra{\psi_q^{(-)}}\hat U_S^\dagger \hat S_A \hat U_S
\ket{\psi_q^{(-)}} ,
\end{equation}
which is equivalent to keeping the interaction-side representation fixed but replacing the source by $\hat S_B\equiv \hat U_S^\dagger \hat S_A\hat U_S$. 
We now characterize the induced source kernel that represents this mismatch in the original interaction-side representation. For the rank-one transformations considered in Ref.~\cite{Epelbaum:2025aan},
\begin{equation}
\hat U_S=1-2\ket{g}\bra{g},\,
\langle g|g\rangle=1 ,
\end{equation}
one has $\hat U_S=\hat U_S^\dagger$ and $\hat U_S^2=1$. The induced source can then be expanded as
\begin{equation}
\begin{aligned}
\hat S_B
&=
\hat U_S^\dagger \hat S_A\hat U_S
\\
&=
\hat S_A
-2\ket{g}\bra{g}\hat S_A
-2\hat S_A\ket{g}\bra{g}
+4\ket{g}\bra{g}\hat S_A\ket{g}\bra{g} .
\end{aligned}
\label{Eq.SM.OperatorExpandedSource}
\end{equation}

The relative coordinate-space matrix elements are therefore
\begin{equation}\label{Eq.SM.TransformedSourceKernelExplicit}
\begin{aligned}
\bra{\mathbf r}\hat S_B\ket{\mathbf r'}
&=
S_0(r)\delta^{(3)}(\mathbf r-\mathbf r')
-2g(\mathbf r)g^*(\mathbf r')
\left[
S_0(r)+S_0(r')
\right]
+4g(\mathbf r)g^*(\mathbf r')
\int \mathrm{d}^3\rho
S_0(\boldsymbol\rho)|g(\boldsymbol\rho)|^2\\
&\equiv
S_0(r)\delta^{(3)}(\mathbf r-\mathbf r')
+\Delta S_U(\mathbf r,\mathbf r').
\end{aligned}
\end{equation}

The first term in Eq.~\eqref{Eq.SM.TransformedSourceKernelExplicit} is the original local source kernel. The remaining term $\Delta S_U(\mathbf r,\mathbf r')$ is the induced generally nonlocal change of the source kernel, including off-diagonal components that appear when the mismatch is represented entirely on the source side. 
In this representation, the mismatch is therefore given by the difference between the induced kernel of $\hat S_B$ and the original local kernel of $\hat S_A$.

At the level of the correlation function, the corresponding change arises entirely from the nonlocal (off-diagonal) kernel structure of $\Delta S_U(\mathbf r,\mathbf r')$,
\begin{equation}
\Delta C_U(q)
\equiv
C_{\rm red}(q)-C_{\rm blue}(q)
=
\int \mathrm{d}^3r\mathrm{d}^3r'
\psi_q^*(\mathbf r)
\Delta S_U(\mathbf r,\mathbf r')
\psi_q(\mathbf r').
\label{Eq.SM.UTCorrectionToCF}
\end{equation}

In this work, the subscripts ``red'' and ``blue'' follow the notation of Fig.~3 of Ref.~\cite{Epelbaum:2025aan}: $C_{\rm red}$ denotes the result obtained with the scattering-equivalent transformed interaction while keeping the original local source fixed, whereas $C_{\rm blue}$ denotes the consistently transformed result, which overlaps with the open-circle reference result. Eq.~\eqref{Eq.SM.UTCorrectionToCF} therefore identifies the red--blue difference with the correlation-level manifestation of the mismatch between the transformed interaction-side representation and the unchanged local source. In the original interaction-side representation, the same difference is encoded in the full matrix element of $\Delta S_U$.

To quantify the induced nonlocality, we introduce the root-mean-square width of the kernel in the relative separation between the two coordinate arguments,
\begin{equation}
\sigma_{\mathbf r-\mathbf r'}
=
\left[
\frac{
\displaystyle
\int
\mathrm{d}^3r\mathrm{d}^3r'
\left|\mathbf r-\mathbf r'\right|^2
\left|\Delta S_U(\mathbf r,\mathbf r')\right|^2
}{
\displaystyle
\int
\mathrm{d}^3r\mathrm{d}^3r'
\left|\Delta S_U(\mathbf r,\mathbf r')\right|^2
}
\right]^{1/2}.
\label{Eq.SM.SigmaOff}
\end{equation}

Thus, $\sigma_{\mathbf r-\mathbf r'}$ characterizes the typical width of the induced kernel in the coordinate difference $\mathbf r-\mathbf r'$. The heat maps of the induced structure $\Delta S_U$ are shown in Fig.~\ref{Fig.InducedNonlocalSource} of the main text.  The left panel gives the Alice reference, for which $\Delta S_U=0$. The middle and right panels show the kernels generated by the two rank-one transformations used for Bob-I and Bob-II, respectively. The quoted values of $\sigma_{\mathbf r-\mathbf r'}$ indicate that the induced source change is not concentrated near the diagonal. In particular, for both transformations, $\sigma_{\mathbf r-\mathbf r'}$ is significantly larger than the source radius $r_0=1.5~{\rm fm}$, showing that the change is broadly distributed in the off-diagonal region. The induced source-kernel change $\Delta S_U$ is therefore not confined to a narrow neighborhood of the diagonal.

The result therefore provides an explicit example in which combining a transformed interaction-side representation with the unchanged local source produces a sizable correlation-level effect. This effect should not be interpreted as an intrinsic shortcoming of femtoscopy; it is the observable consequence of an uncontrolled source--interaction mismatch.

This raises the practical question of how a consistent
source--interaction pairing can be established when no common
microscopic derivation is available.

\section{An operational source--interaction convention}
\label{App:Source_convention}

We therefore introduce an operational source--interaction convention: a prescription for fixing the source-side representation and establishing a compatible interaction-side representation.

To fix the source-side convention, we take the reduced density operator produced in a chosen reference heavy-ion collision system, denoted by $\hat\rho_{\rm ref}$, as the reference source. We then choose a basis
\begin{equation}
\mathcal B_S
\equiv
\mathcal B_{\rm ref}
=
\left\{
|\mathbf r_{\rm ref}\rangle
\right\},
\end{equation}
in which the matrix representation of $\hat\rho_{\rm ref}$ is approximately diagonal at the spatial resolution relevant to the correlation measurement.

A pointer basis dynamically selected via decoherence provides a possible physical motivation for this choice. We do not, however, identify $\mathcal B_{\rm ref}$ with a unique pointer basis. There may be several bases in which the reference density matrix is approximately diagonal to the accuracy relevant for the observable. The source convention selects one such basis and uses it consistently.

In this basis, the reference source satisfies
\begin{equation}
\langle\mathbf r_{\rm ref}|
\hat\rho_{\rm ref}
|\mathbf r'_{\rm ref}\rangle
\simeq
S_{\rm ref}(\mathbf r)
\delta^{(3)}(\mathbf r-\mathbf r')
\end{equation}
or, more generally, is strongly concentrated near $\mathbf r=\mathbf r'$. The convention fixes the basis in which the source is represented and in which the diagonal approximation is imposed. It neither identifies the abstract density operator with a particular basis-dependent kernel nor requires density operators produced by other mechanisms to be approximately diagonal in the same basis. Once $\mathcal B_{\rm ref}$ has been fixed, any other physical source is represented in that basis and may contain both a different diagonal distribution and nonvanishing off-diagonal components.

With this source basis fixed, the on-shell scattering observables of a reference pair $A$ first constrain the relevant scattering-equivalence class, while fitting its measured correlation function operationally calibrates a compatible interaction-side realization within that class. This establishes an effective source--interaction pairing within the chosen framework.
An exact microscopic match cannot generally be demonstrated. Here, the pairing is considered effective as long as any residual mismatch changes the correlation function by less than the relevant experimental and theoretical accuracies over the kinematic range of interest.

Weak sensitivity to extremely short-range structures is not required for the calibration itself, but it provides additional robustness against such residual mismatch.
The impact of a microscopic mismatch on the correlation function is determined by its contribution to the full source-weighted integral, rather than by the magnitude of the microscopic difference alone. In the systems studied in Refs.~\cite{Gobel:2025afq,Molina:2025lzw}, realistic off-shell or interaction-model variations were found to change the correlation function by about $6\%$ or less. The ALICE analyses further show that the relative contribution of the sub-femtometer region to the source-weighted correlation integral is limited, where differences among wave functions associated with different interaction-side representations are expected to be most pronounced~\cite{ALICE:2023sjd,ALICE:2026tqh}.

Within the present framework, CF-to-CF predictions are well defined and practically feasible. 
Once a theoretical model has been calibrated against the measured correlation function of the reference pair $A$, interactions for another pair $B$ derived consistently from the same model inherit the calibrated interaction-side convention and can therefore be combined with the same source-side convention for CF-to-CF predictions.
In an EFT, for example, fitting the relevant low-energy constants calibrates the EFT within the chosen Lagrangian formulation, and interactions in other channels derived from the same EFT inherit this convention.

For a different production mechanism, the source density operator itself changes, while the reference basis remains fixed. The new source is represented as
\begin{equation}
\rho_{\rm new}(\mathbf r,\mathbf r')
=
\langle\mathbf r_{\rm ref}|
\hat\rho_{\rm new}
|\mathbf r'_{\rm ref}\rangle .
\end{equation}
If its coherence properties differ from those of the reference source, this kernel may exhibit a different diagonal distribution and nonvanishing off-diagonal elements. Since both sources are represented in the same fixed reference basis, differences between their kernels reflect physical differences between the underlying source density operators.

Returning to the example of Ref.~\cite{Epelbaum:2025aan}, the mismatch produced by combining the transformed interaction-side representation with the unchanged local source lies beyond the validity range of the calibrated effective pairing.

\section{Coupled-channel quantum maps and channel-projected correlation functions}
\label{App:coupled_channel_maps}

In this section, we show how the density operator formulation can be extended to a retained coupled-channel subsystem. The purpose is not to construct a full coupled-channel scattering theory, but to clarify how channel structure, open-system evolution, and channel-projected measurements can be described within the same operator language used in the main text.

We adopt the usual physical-channel convention used in scattering theory and femtoscopy. A channel denotes a physical sector specified by the asymptotic particle content and the relevant internal quantum numbers, such as particle species, particle number, thresholds, spin, or other discrete quantum numbers. Continuous kinematic variables, such as total momentum, relative momentum, and energy, together with angular-momentum labels, are treated as basis labels inside a given channel. With this convention, momentum or energy redistribution induced by the environment does not by itself imply a transition to a different physical channel.

It is useful to distinguish between two independent classifications. The first is whether the retained observed subsystem contains a single physical channel or multiple coupled channels. The second is whether, after tracing over the environment, the retained subsystem evolves as an effectively closed system or continues to interact with the environment as an open system. These possibilities are summarized in Table~\ref{tab:open_closed_coupled_single}.

\begin{table}[htbp]
\centering
\caption{Single-channel/coupled-channel and closed/open descriptions.}
\label{tab:open_closed_coupled_single}
\begin{tabular}{p{0.20\textwidth}p{0.47\textwidth}p{0.25\textwidth}}
\hline
Case & Physical meaning & Evolution form \\
\hline
Closed + single channel &
Only one physical channel is retained, and after the tracing stage, the subsystem no longer interacts significantly with the environment. &
$\hat\rho\to \hat U\hat\rho\hat U^\dagger$ \\

Open + single channel &
Only one physical channel is retained, but the subsystem still interacts with the environment. The environment may modify momentum, energy, phase, or coherence without changing the channel identity. &
$\hat\rho\to \mathcal E[\hat\rho]=\sum_a \hat K_a\hat\rho \hat K_a^\dagger$ \\

Closed + coupled channels &
Several physical channels are retained, and after the tracing stage, the retained coupled-channel subsystem evolves unitarily. &
$\boldsymbol\rho\to \boldsymbol U\boldsymbol\rho\boldsymbol U^\dagger$ \\

Open + coupled channels &
Several physical channels are retained, and after the tracing stage, the retained subsystem continues to interact with the environment. &
$\boldsymbol\rho\to
\boldsymbol{\mathcal E}[\boldsymbol\rho]
=\sum_a \boldsymbol K_a\boldsymbol\rho\boldsymbol K_a^\dagger$ \\
\hline
\end{tabular}
\end{table}

Table~\ref{tab:open_closed_coupled_single} emphasizes that open-system dynamics and coupled-channel dynamics are not the same distinction. A system may be open while retaining only one physical channel. Conversely, a retained coupled-channel subsystem may still be closed if its post-tracing evolution is unitary.

Let the retained coupled-channel Hilbert space be written as
\begin{equation}
\mathcal H_{\rm cc}
=
\bigoplus_{\alpha=1}^{N_{\rm ch}}\mathcal H_\alpha ,
\end{equation}
where $\alpha$ labels the retained physical channels. If the full system is decomposed as
\begin{equation}
\mathcal H_{\rm tot}
=
\mathcal H_{\rm cc}\otimes\mathcal H_E ,
\end{equation}
then tracing over the environmental degrees of freedom gives the coupled-channel reduced density operator
\begin{equation}
\boldsymbol\rho_{\rm in}
=
\mathrm{Tr}_{\mathcal H_E}
\hat\rho^{\rm tot}.
\end{equation}
The boldface notation indicates that the density operator acts on the full retained coupled-channel space. It has the channel-block structure
\begin{equation}
\boldsymbol\rho
=
(\hat\rho_{\alpha\beta}) .
\end{equation}
The diagonal block $\hat\rho_{\alpha\alpha}$ describes the density operator associated with channel $\alpha$, while the off-diagonal block $\hat\rho_{\alpha\beta}$ with $\alpha\neq\beta$ represents coherence between different retained channels.

Within the CPTP description adopted here, the open-system evolution of the retained coupled-channel subsystem can be written as a quantum map,
\begin{equation}
\boldsymbol\rho'
=
\boldsymbol{\mathcal E}[\boldsymbol\rho]
=
\sum_a
\boldsymbol K_a
\boldsymbol\rho
\boldsymbol K_a^\dagger .
\end{equation}
Each Kraus operator $\boldsymbol K_a$ also has channel blocks. We denote these blocks by $\hat K_{a,\alpha\beta}$, where the first index labels the output channel and the second index labels the input channel. Thus, $\hat K_{a,\alpha\beta}$ maps states from channel $\beta$ into channel $\alpha$.

The corresponding block evolution is
\begin{equation}
\hat\rho'_{\alpha\beta}
=
\sum_a
\sum_{\mu,\nu}
\hat K_{a,\alpha\mu}
\hat\rho_{\mu\nu}
\hat K_{a,\beta\nu}^\dagger .
\label{Eq.CCBlockEvolution}
\end{equation}
This relation shows that the output $(\alpha,\beta)$ block is not determined only by the input $(\alpha,\beta)$ block. In general, all input channel blocks $\hat\rho_{\mu\nu}$ can contribute through the corresponding Kraus blocks. In particular, if the final observed channel is $\lambda$, the relevant output block is
\begin{equation}
\hat\rho'_{\lambda\lambda}
=
\sum_a
\sum_{\mu,\nu}
\hat K_{a,\lambda\mu}
\hat\rho_{\mu\nu}
\hat K_{a,\lambda\nu}^\dagger .
\label{Eq.CCObservedBlock}
\end{equation}

We now introduce the measurement operator for a selected observed channel. If the final measurement is restricted to channel $\lambda$, the measurement operator acts only inside that channel and is denoted by $\hat M_\lambda$. Embedded into the full coupled-channel space, it becomes a block operator $\boldsymbol M_\lambda$ with only one nonzero block,
\begin{equation}
(\boldsymbol M_\lambda)_{\alpha\beta}
=
\delta_{\alpha\lambda}
\delta_{\beta\lambda}
\hat M_\lambda .
\end{equation}
The channel-projected yield, or unnormalized correlation function, can therefore be written as a trace over the retained coupled-channel space,
\begin{equation}
C_\lambda
=
\mathrm{Tr}_{\rm cc}
\left[
\boldsymbol M_\lambda
\boldsymbol\rho'
\right].
\end{equation}
Since $\boldsymbol M_\lambda$ only has a nonzero $(\lambda,\lambda)$ block, this is equivalent to
\begin{equation}
C_\lambda
=
\mathrm{Tr}_{\mathcal H_\lambda}
\left[
\hat M_\lambda
\hat\rho'_{\lambda\lambda}
\right].
\end{equation}
Substituting Eq.~\eqref{Eq.CCObservedBlock}, one obtains
\begin{equation}
C_\lambda
=
\sum_a
\sum_{\mu,\nu}
\mathrm{Tr}_{\mathcal H_\lambda}
\left[
\hat M_\lambda
\hat K_{a,\lambda\mu}
\hat\rho_{\mu\nu}
\hat K_{a,\lambda\nu}^{\dagger}
\right].
\label{Eq.CCProjectedYield}
\end{equation}
This is the channel-projected density-operator expression. It shows that the yield observed in the final channel $\lambda$ does not necessarily originate only from the initial $\lambda\lambda$ block. All retained channel blocks that can be dynamically mapped into the observed channel may contribute.

Several useful limits follow directly from this expression. If the retained coupled-channel subsystem is closed after the tracing stage, the quantum map reduces to a unitary coupled-channel evolution. If only one channel is retained and the subsystem is closed, the formula reduces to the single-channel closed-system form used in the sudden approximation in the main text. If only one channel is retained but the subsystem remains coupled to the environment, the result is a single-channel open-system evolution rather than a coupled-channel evolution.

This distinction is also useful for interpreting absorption. Within the CPTP description adopted here, the evolution of the full retained coupled-channel space is trace preserving. The probability weight of one selected observed channel need not be conserved, however, because probability can flow into other retained channels. In this sense, absorption means loss from the selected observed channel, not loss of total probability in the full retained space.

\end{document}